\documentclass[twocolumn,
superscriptaddress,
nofootinbib,
 amsmath,amssymb,
 aps,
floatfix
]{revtex4-1}

\usepackage{times}

\usepackage{amsfonts}
\usepackage{amssymb}
\usepackage{graphicx}
\usepackage{bbold}
\usepackage{dsfont}

\usepackage{accents}
\usepackage{graphicx}
\usepackage{gensymb} 
\usepackage{dcolumn}
\usepackage{bm}
\usepackage{amsmath}
\usepackage{graphicx}
\usepackage{yfonts}
\usepackage{float}
\usepackage{mathtools}
\usepackage{comment}

\DeclareMathOperator{\tr}{Tr}
\newcommand{\id}{\mathds{1}}
\usepackage{amsthm}

\usepackage{notes2bib}
\bibnotesetup{
note-name = ,
use-sort-key = false
}

\newcommand{\ket}[1]{\left| #1 \right\rangle}

\newcommand{\ketbra}[2]{\left|#1\middle\rangle\middle\langle#2\right|}

\newcommand{\Bra}[1]{{ \langle \! \langle{#1}\vert }}
\newcommand{\Ket}[1]{{ \vert {#1}  \rangle \!  \rangle}}
\newcommand{\KetBra}[1]{{\Ket{#1}\!\Bra{#1} }}

\usepackage[normalem]{ulem}
\usepackage{xcolor}

\usepackage {hyperref}
\hypersetup{
     colorlinks   = true,
     linkcolor    = blue,
     citecolor    = blue,
     urlcolor     = blue,
     }
\makeatother

\begin{document} 
\preprint{APS/123-QED}


\title{Experimental characterisation of a non-Markovian quantum process}

\author{K. Goswami } 
\email{k.goswami@uq.edu.au}
\affiliation{Centre for Engineered Quantum Systems, School of Mathematics and Physics, University of Queensland, QLD 4072 Australia}
\author{C. Giarmatzi}
\affiliation{University of Technology Sydney, Centre for Quantum Software and
Information, Ultimo NSW 2007, Australia }
\author{C. Monterola}
\affiliation{Aboitiz School of Innovation, Technology, and Entrepreneurship, Asian Institute of Management, Makati  Metro Manila 1229, Philippines.}
\author{S. Shrapnel}
\affiliation{Centre for Engineered Quantum Systems, School of Mathematics and Physics, University of Queensland, QLD 4072 Australia}
\author{J. Romero}
\email{m.romero@uq.edu.au}
\affiliation{Centre for Engineered Quantum Systems, School of Mathematics and Physics, University of Queensland, QLD 4072 Australia}
\author{F. Costa}
\email{f.costa@uq.edu.au}
\affiliation{Centre for Engineered Quantum Systems, School of Mathematics and Physics, University of Queensland, QLD 4072 Australia}

\begin{abstract}
Every quantum system is coupled to an environment. Such system-environment interaction leads to temporal correlation between quantum operations at different times, resulting in non-Markovian noise. In principle, a full characterisation of non-Markovian noise requires tomography of a multi-time processes matrix, which is both computationally and experimentally demanding. In this paper, we propose a more efficient solution.  We employ machine learning models to estimate the amount of non-Markovianity, as quantified by an information-theoretic measure, with tomographically incomplete measurement. We test our model on a quantum optical experiment, and we are able to predict the non-Markovianity measure with $90\%$ accuracy. Our experiment paves the way for efficient detection of non-Markovian noise appearing in large scale quantum computers.
\end{abstract}

\maketitle
\section{Introduction}

One of the biggest challenges in emerging quantum technologies is the efficient characterisation of noise, which originates from the unavoidable interaction of a system of interest with the surrounding environment \cite{Wiseman2009}. A particularly problematic case is that of non-Markovian noise, wherein the environment retains a memory of its past interactions with the system, leading to correlations in the system’s evolution at different times.

Most noise characterisation methods, such as randomised benchmarking \cite{Emerson2005, Knill2008}, rely on the assumption of Markovianity \cite{Epstein2014, Ball2016}. However, non-Markovian noise is already prominent in present-day quantum devices \cite{morris2019nonmarkovian, White2020}. Therefore, it is important to find efficient methods to estimate the amount of non-Markovianity in a quantum process \cite{LI20181}.

Recently, an operational approach has been proposed, based on the ``combs'' \cite{chiribella09b} or ``process matrix'' \cite{oreshkov12} formalism, which overcomes previous theoretical difficulties and allows, in principle, for a complete characterisation of non-Markovian dynamics from operations and measurements on the system alone \cite{pollock_operational_markov, giarmatzi2018witnessing}. However, this method relies on complete tomography of a multi-time process \cite{costa2016, Pollock_pra}, which requires measuring an exponential number of multi-time correlations. 
In Ref.~\cite{Shrapnel2018}, it was shown that one can successfully train a machine learning model to estimate a measure of non-Markovianity, without full process tomography. This work, however, used only numerically simulated data, and was not tested in an experimental setting.

Here, we use a quantum-optics experimental setup to implement a non-Markovian process---specifically, a process with initial classical correlations between system and environment. We encode quantum states in the polarisation of photons and apply unitary transformations using wave-plates. We introduce non-Markovian noise through correlated random unitaries, performed before and after a probe unitary. Our data comprises the Stokes parameters, obtained through a final measurement, conditional on choosing the probe unitary from a set of three. We train a suite of different supervised machine learning models to predict non-Markovianity---as quantified by an entropic measure introduced in Ref.~\cite{pollock_operational_markov}. 

Our method achieves a high accuracy in the estimation of non-Markovianity, even though the training data is far from being tomographically complete. The best results were achieved by a quadratic regression model ($R^2$ of $0.89$ and Mean Absolute Error (MAE) of $0.045$).
Our work expands on the growing literature of machine-learning methods~\cite{Niu2019, Luchnikov2019, Luchnikov2020, Fanchini2020, Guo2020} and on the experimental characterisation of quantum non-Markovianity~\cite{Liu2011, Chiuri2012, Ringbauer2015, Yu2018, Xiong2019, Wu2020, Silva2020}.

We present the work following way. In Section~\ref{sec:2}, we introduce the framework of the process matrix, the measure of non-Markovian noise, and procedure of our data acquisition. In Section~\ref{sec:3}, we describe our experiment. In Section~\ref{sec:4}, we analyse our experimental data using polynomial regression and present our results. In the Appendix, aside from polynomial regression on the experimental data, we present our results on the simulated data and our results obtained by other machine learning algorithms.

\section{Theory}\label{sec:2}

\subsection{Formulation of quantum processes}

Non-Markovian quantum processes are often described in terms of dynamical maps representing the evolution of the system's reduced state \cite{Rivas2014}. However, such a description does not capture multi-time correlations mediated by the environment and can fail entirely in the presence of initial system-environment correlations \cite{Pechukas1994, stelmachovic2001}. Here we use instead the process matrix formalism~\cite{oreshkov12,oreshkov15}, following a recent approach~\cite{pollock_operational_markov, Pollock_pra} that has reformulated in operational terms the theory of quantum stochastic processes \cite{Lindblad1979, Accardi82}.
We consider a scenario where a system of interest undergoes a sequence of  arbitrary operations (such as unitaries or measurements) at well-defined instants of time. Let us label $A, B,\dots$ the times at which the operations are performed (we can think of these labels as referring to ``measurement stations''). The most general operation, say at $A$, is described by a
Completely Positive (CP) map $\mathcal{M}^{A_I\rightarrow A_O}$ that maps the input system of the operation $A_I$ to its output system $A_O$. The set of all measurement outcomes corresponds to a \emph{quantum instrument}~\cite{davies70}, namely a collection of CP maps $\mathcal{J}^A = \{\mathcal{M}^A\}$ that sum up to a CP and Trace Preserving (CPTP) map. Note that, as a particular case, the instrument can contain a single map, representing a deterministic operation with no associated measurement (for example, a unitary transformation). Also, we typically take the last output system to be trivial (as the system is discarded afterwards), in which case the instrument reduces to a Positive Operator Valued Measure (POVM).

In a given quantum process, the joint probability for outcomes to occur at measurement stations $A, B, ...$ (corresponding to CP maps $\mathcal{M}^A, \mathcal{M}^B, \cdots $) is given by
\begin{align}
    p({\cal M}^{A}, {\cal M}^{B},\cdots | {\cal J}^{A}, {\cal J}^{B}, \cdots) = \nonumber \\ 
    \tr[W^{A_IA_OB_IB_O\cdots} (M^{A_IA_O} \otimes M^{B_IB_O} \otimes \cdots)],
\label{eq:prob}
\end{align}
where $M^{A_IA_O}, M^{B_IB_O}, \cdots$ are the \emph{Choi matrices}~\cite{Choi1975, choi75b} of the corresponding maps and $W^{A_IA_OB_IB_O\cdots}$ is the process matrix that surrounds the measurement stations $A, B, \cdots$ and lives on the Hilbert space of their combined inputs and outputs. A Choi matrix, say $M^{A_IA_O}\in {\cal L}({\cal H}^{A_I}\otimes{\cal H}^{A_O})$, that is isomorphic to a CP map ${\cal M}^{A} : {\cal L}({\cal H}^{A_I}) \rightarrow {\cal L}({\cal H}^{A_O})$, is defined as $M^{A_IA_O} := [{\cal I} \otimes {\cal M}(\KetBra{\id})]^T$. $\cal I$ is the identity map, $\Ket{\id} = \sum_{j=1}^{d_{A_I}}\ket{jj} \in {\cal H}^{A_I} \otimes {\cal H}^{A_I}$, $\{\ket{j}\}^{d_{A_I}}_{j=1}$ is an orthonormal basis on ${\cal H}^{A_I}$ and ${T}$ denotes matrix transposition in that basis and some basis of ${\cal H}^{A_O}$. The process matrix $W$ is also known as process tensor \cite{pollock_operational_markov}, or comb \cite{chiribella08}, and it is equivalent to a quantum channel with memory \cite{Kretschmann2005}.

In this formalism, it was found that the process matrix of a Markovian process should have the following form~\cite{costa2016, Pollock:2018ab, Giarmatzi:2018aa, giarmatzi2018witnessing}
\begin{align}
    W^{AB\cdots}_{\textup {M}} = \rho^{A_I}\otimes T^{A_OB_I}\cdots,
\label{eq:Wab_m}
\end{align}
where $\rho$ is the density matrix of the initial state and by $T^{A_OB_I}$ we denote the Choi matrix of the channel $\mathcal{T}^{A\rightarrow B}$, defined as above but without the transposition---the same applies throughout the paper to all the Choi matrices of channels in a process matrix.

The form of a Markovian process matrix in Eq.~\eqref{eq:Wab_m} has a straightforward interpretation: just before the first operation (measurement station $A$), the system is in the initial state $\rho$. Between the first and second operation, the system evolves according to a CPTP map $\mathcal{T}$, which is uncorrelated with the initial state, and so on, with all evolutions independent of each other and of the initial state. Conversely, any process matrix that cannot be expressed in such a product form represents non-Markovian evolution, where the environment mediates correlations between the initial state and subsequent evolutions. To determine whether a process is Markovian, one needs first to reconstruct the process matrix from experimental data through process tomography---which generally involves non-destructive measurements at each station \cite{costa2016}---and then check if it $W$ can be written in the product form \cite{Giarmatzi:2018aa}. In the following, we provide a method to detect non-Markovianity without having the full process matrix---instead, with incomplete data about the process, we can estimate with high accuracy a measure of non-Markovianity.

\subsection{Our non-Markovian process}

\begin{figure}[!h]
\begin{center}
\includegraphics[width=\columnwidth]{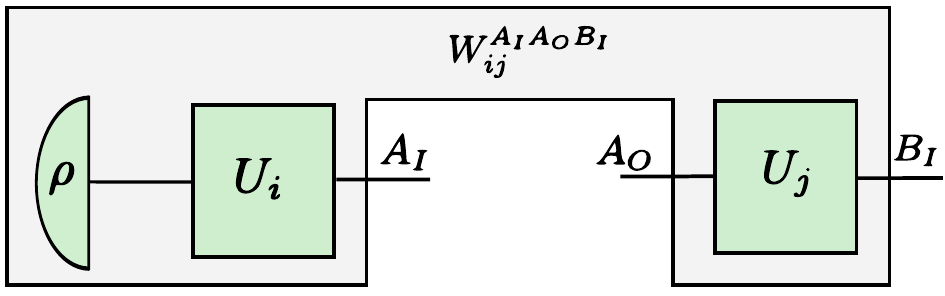}
\caption{ A process based on a specific instance of unitary operations $U_i $ and $U_j $. The pair of unitaries occurs with probability $p(i,j)$. This makes our overall process $W$ a convex combination of the constituent processes, $W_{ij}$, i.e., $W{=}\sum_{i,j}p(i,j)W_{ij}$. This process $W$, operationally, represents the environment. The experimenalist accesses the open slot $A_IA_O$ with a probe unitary $U_K$, as in Eq.~\eqref{Eq:middle_unitary} and $B_I$ with a Pauli measurement.}
\vspace{-3mm}
\label{fig:process}
\end{center}
\end{figure}

We experimentally implement a non-Markovian quantum process with memory.  We implement a process with only two ``stations'', $A$ and $B$, and where the initial state is classically correlated with the evolution from $A$ to $B$. This is a particular case of a non-Markovian process with classical memory~\cite{giarmatzi2018witnessing}.
We do this in two steps. We start with some initial state $\rho$ followed by two operations $U_i, U_j$. The operations are unitaries from the Pauli group, $U_i \in \{ \sigma_i, i = \{0,1,2,3\}\}$ and $U_j \in \{ \sigma_j, j = \{0,1,2,3\}\}$, where $\sigma_0 = \id, \sigma_1 = X, \sigma_2 = Y, \sigma_3 = Z$. We insert $A$ between $U_i$ and $U_j$ and $B$ after $U_j$ (Fig. \ref{fig:process}). In this first step, for a given pair of unitaries $(U_i,U_j)$, we obtain the following Markovian process
\begin{align}
W_{ij}^{A_IA_OB_I} = (\sigma_i\rho \sigma_i ^\dagger)^{A_I} \otimes [[\sigma_j]]^{A_OB_I}.
\label{Eq:constitent_process}
\end{align}

In the second step, we simulate a non-Markovian environment by introducing correlations between the initial state and the unitary. This is done by sampling the processes $W_{ij}$ according to some probability distribution $p(i,j)$. The resulting process matrix has the form
\begin{align}
 W^{A_IA_OB_I} = \sum_{i,j} p(i,j) W_{ij}^{A_IA_OB_I}.  
 \label{Eq:convex_combination_process}
\end{align}

To obtain processes with a varying degree of non-Markovianity, the distribution of the weights $p(i,j)$ is chosen according to the discrete random variables $I$ and $J$ governed by the joint probability mass function (pmf) $p(i,j){:=}p(I{=}i,J{=}j)$. From Eq.~\eqref{Eq:convex_combination_process}, it is clear that when the random variables $I$ and $J$ are independent, the overall process reduces to the product form of Eq.~\eqref{Eq:constitent_process} and hence it becomes a Markovian process. To capture the non-Markovian effect, we model the joint probability $p(i,j)$ as 
\begin{align}
 p(i,j){=}p(i)\left[q\delta_{ij}+(1-q)p(j)\right],  \label{Eq:joint_probability}
\end{align}
Here $q{\in}[0,1]$ denotes the strength of correlation between the random variables $I$ and $J$ with $q{=}0$ being mutually independent events and $q{=}1$ being the maximum correlation, i.e. $\sigma_j {=} \sigma_i$. We assume that the marginal probabilities $p(i)$ and $p(j)$ to be the same probability mass functions. To define the probability mass function, for $p(i{=}0)$, we chose a random number uniformly distributed between 0 and $R{\geq}1$. For the remaining $p(i{\neq} 0)$, we chose random numbers uniformly distributed between 0 and 1. We normalise the random numbers at the end to form a valid probability mass function. A high value of $R$ signifies the evolution is less prone to error, i.e. the corresponding random operation is biased towards identity. Note that the process becomes Markovian with either $q{=}0$ or $R{\rightarrow}\infty$.

As a measure of non-Markovianity we use the quantum relative entropy~\cite{Wilde13, pollock_operational_markov, morris2019nonmarkovian, giarmatzi2018witnessing} between the process and the associated Markovian one:
\begin{align}
    S(\tilde{W}||\tilde{W}_{\mathrm{Markov}}) := \tr[\tilde{W}\cdot(\log \tilde{W} {-}\log \tilde{W}_{\mathrm{Markov}})],
    \label{Eq:non-Markovianity_measure}
\end{align}
where $\tilde{W}_{\mathrm{Markov}} := \tr_{A_OB_I}{\tilde{W}}\otimes\tr_{A_I}{\tilde{W}}$ and $\tilde{W}:=W/2$ is the process matrix normalised to have unit trace (obtained dividing the original process matrix by the dimension of the output system, $A_O$ in this case).

\begin{figure}[!h]
\begin{center}
\includegraphics[width=\columnwidth]{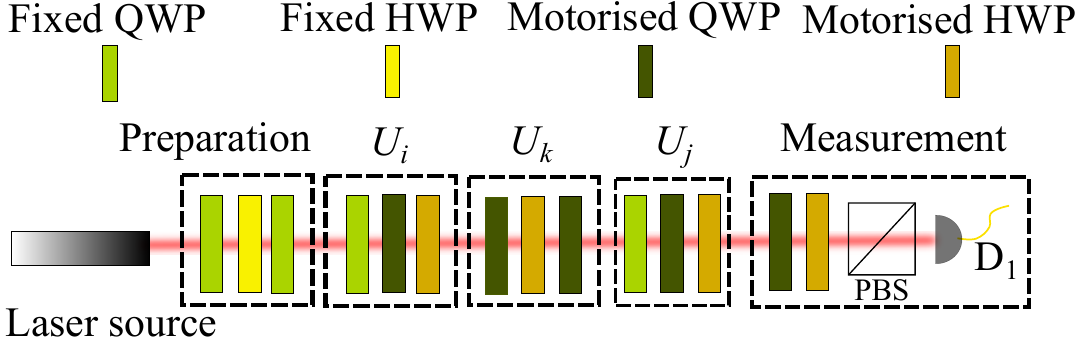}
\caption{Experimental setup. We use polarisation of light to encode the quantum state. The experiment is divided into five stages --- the first stage is state preparation, the second one implements the unitary $U_i$, the third stage represents unitary $U_k$, the fourth one denotes $U_j$ and finally the last stage denotes polarisation measurements.}
\vspace{-3mm}
\label{fig:Experimental_setup}
\end{center}
\end{figure}

In each realisation of $W_{ij}^{A_IA_OB_I}$ with a pair of unitaries $U_i$ and $U_j$, we insert at $A$ a unitary operation $U_k$ and at $B$ we perform state tomography. Each such process $W_{ij}^{A_IA_OB_I}$ has a circuit representation as shown in Figure~\ref{fig:process} and an experimental realisation as shown in Fig~\ref{fig:Experimental_setup}. The unitary operations of $A$ are a set of rotated Pauli operations
\begin{align}
    U_k{=}R_{\hat{n}}(\alpha)\sigma_kR_{\hat{n}}(\alpha)^\dagger,
    \label{Eq:middle_unitary}
\end{align}
where $k{=}\{0,1,2\}$ and $R_{\hat{n}}(\alpha)$ denotes a rotation by $\alpha$, around an arbitrary axis $\hat{n}$ in the Bloch sphere, given by
\begin{align}
    & R_{\hat{n}}(\alpha) = \cos{\frac{\alpha}{2}}\mathbb{1} - i \sin{\frac{\alpha}{2}}(\hat{n}.\Vec{\sigma}) \label{Eq:rotation}, \\
    & \hat{n}.\Vec{\sigma} = \sin{\beta}\sin{\gamma}\sigma_1+\cos{\beta}\sin{\gamma}\sigma_2+\cos{\gamma}\sigma_3 \label{Eq:arbit_axis}.
\end{align}

Briefly, the experimental procedure of realising a process with classical memory and taking data consists of the following steps: (1) Choosing a pair of variables $(q,R)$ to obtain the weights $p(i,j)$, (2) Realising the processes $W_{ij}$, and for each one, taking data $D_{ij}$ by running through the operations at $A$ and $B$, and (3) Calculating the data $D(q,R) = \sum_{ij} p(i,j) D_{ij} $. This final data is our input to a model that predicts the non-Markovianity of the process $ W = \sum_{i,j} p(i,j) W_{ij}$.

To complete the set of training, validation, and test data for our model, we calculated the non-Markovianity for the realised processes--- the label for each data $D(q,R)$. For that, we need the explicit description of the realised process matrix, which we can obtain from the above theoretical description.

We stress here that the input to the model that predicts the amount of non-Markovianity is data taken by inserting the operations $A$ and $B$ into the process. These provide incomplete information about the process. The full information would be provided by informationally complete operations, for example, a prepare-and-measure operation at $A$, and state tomography at $B$ (with a minimum of $64$ operations for a $3$-qubit $W$, such as ours). In our case, while $B$ performs state tomography, $A$ performs 3 Pauli unitary operations. However, even with this incomplete information, the model is able to predict the chosen measure of non-Markovianity with $\approx 90\%$ accuracy.

\subsection{Generating data and labels} 

One key point to consider in any predictive modelling is to avoid inherent bias in the training dataset. This bias can be manifested in terms of trivial transformation of the initial state. We account for this by choosing a suitable initial state, $\rho$, that leads to processes resulting non-trivial output data. Our choice of state is $\rho{=}\ketbra{\psi}{\psi}$, with $\ket{\psi}{=}0.16\ket{0}{+}0.99e^{-i.0.16\pi}\ket{1}$. To model the probability mass function as in Eq.~\eqref{Eq:joint_probability}, we take the 10 pairs of $q$ and $R$  listed in Table~\ref{Table:q_R_pairs}.

\begin{table}
\begin{tabular}{|c  | c|}
 \hline

 $\qquad  q \qquad $ & $\qquad  R \qquad  $ \\

 \hline 

 $0.8$ & 1 \\
  
$0.8$ & 1.5 \\
 
$0.8$ & 1.25 \\
 
$0.9$ & 1 \\

$0.9$ & 1.5 \\

$0.9$ & 1.25 \\

$0.95$ & 1 \\

$0.95$ & 1.5 \\

$0.95$ & 1.25 \\

$1$ & 1 \\
\hline 
\end{tabular}
 \caption{Pairs of $q$ and $R$ to model our joint pmf as defined in Eq.~ \eqref{Eq:joint_probability}.  For each pair of $q$ and $R$ we generate 100 pmfs, thus for 10 pairs we have a total of 1000 datasets.}
\label{Table:q_R_pairs} 
 \end{table}

For each pair, we generate 100 joint probability mass functions thus creating 1000 different processes as in Eq.~\eqref{Eq:convex_combination_process} which are then divided into 100 groups classified by a given pair of $q$ and $R$. Note that a specific instance of the experiment corresponds to a pair of unitaries sampled randomly from the underlying pmf. To experimentally realise the process in Eq.~\eqref{Eq:convex_combination_process} described by a particular pmf, we need to perform repeated trials. In our experiment, we take 50 samples of each pmf. This finite sampling yields an experimentally realised process $W_{\mathrm{exp}}$ defined as 
\begin{align}
W_{\mathrm{exp}} {=} \sum_{i,j}\tilde{p}(i,j)W{ij}.
\end{align}
Here, $\tilde{p}(i,j)$ is the frequency of occurrence of the particular unitary pair $(U_i, U_j)$, and $W_{ij}$ is the constituent process defined in Eq.~\eqref{Eq:constitent_process}. For each $(U_i, U_j)$, we apply unitary operation $U_k$ at the second time-step as defined in Eq.~\eqref{Eq:middle_unitary} with $\alpha{=}\beta{=}\gamma{=}\pi/8$. As discussed earlier, we interpret $U_k$ as an experimentally-controlled intervention, while $W_{\mathrm{exp}}$ simulates a noisy environment. Thus, in each instance, we have the state evolving through an overall unitary operation $U_jU_kU_i$. We measure the output state in the Pauli basis. Taking average over $U_i$ and $U_j$, we get the mixed state $\rho_k{=}\sum_{i,j}\tilde{p}(i,j)(U_jU_kU_i)\rho_{\mathrm{in}}(U_i ^\dagger U_k ^\dagger U_i ^\dagger)$. This state, when measured in $\sigma_l$ basis, yields a Stokes parameter $S_{lk}$ where
\begin{align}
    S_{lk} & = \tr(\sigma_l\rho_k) \nonumber \\ 
    & = \tr\bigg\{\bigg(([[U_k^*]])^{A_IB_I}{\otimes}\sigma_l^{B_O}\bigg)\cdot W_{\mathrm{exp}}^{A_IA_OB_I}\bigg\}.
\end{align}
Note that both $k$, $l\in\{0,1,2\}$. For each process $W_\mathrm{exp}$, we have total of 9 Stokes parameters---from now on we refer to them as \emph{datapoints}. We evaluate the measure of the non-Markovianity associated with the process $W_\mathrm{exp}$ using Eq.~\eqref{Eq:non-Markovianity_measure} with $W=W_{\mathrm{exp}}$---from now on, we refer to these measures as \emph{labels}. Thus, we have a total number of 1000 labeled data, each containing 9 datapoints and the corresponding label.

\section{Experiment} \label{sec:3}
We show the experimental schematic in Fig~\ref{fig:Experimental_setup}. We start with a heavily attenuated laser centred at 820 nm wavelength to create weak coherent states with 10000 counts per second. We encode the state in the photon's polarisation. Our experiment is divided into the following stages: state preparation, implementing the unitaries $U_i$, $U_k$ and $U_j$, and state measurement. The polarisation state is prepared using a series of waveplates (Fig. \ref{fig:Experimental_setup}).  The arbitrary unitaries in polarisation were implemented using three waveplates, a half-waveplate (HWP) in between two quarter-waveplates (QWP) as in Fig. \ref{fig:Experimental_setup} \cite{SIMON_3polarisation}. To automate the transition between unitaries, we used motorised stages. Each $U_i$ and $U_j$ change within the Pauli group. For each of them we need only two motorised stages and a fixed QWP at 0\degree (the angles for the waveplates are given in Table~\ref{Table:unitary_angles}). For the unitary $U_k$, we use three motorised stages. We monitor the motorised stages using a LabVIEW-controlled Newport XPS series motion controller through a TCP/IP protocol and a Newport SMC 100 series motion controller  with a serial communication to a computer. For preparing the state $\ket{\psi}$, we use another series of waveplates. Since the first QWP of $U_i$ is set to a fixed angle at 0\degree, we can absorb that in the state preparation. After successful implementation of state preparation and the unitaries, we measure the Stokes parameter of the output light using a standard setup of QWP-HWP and polarising beamsplitter, as shown in Fig.~\ref{fig:Experimental_setup}.

\begin{table}[h]
\begin{tabular}{|c|c|c|c|}
 \hline
 Unitary & QWP  & QWP & HWP  \\
 \hline 
 $\mathbb{1}{\equiv}\sigma_0$ & 0 & 0 & 0 \\
 \hline 
 $X{\equiv}\sigma_1$ & 0 & $\frac \pi 2$&$\frac \pi 4$ \\
 \hline
 $Y{\equiv}\sigma_2$ & 0 &0 &$\frac \pi 4$ \\
 \hline
 $Z{\equiv}\sigma_3$ & 0 & $\frac \pi 2$& 0\\
\hline 
 \end{tabular}
 \caption{Angles for motorised wave plates to implement $U_i$ and $U_j$.}
 \label{Table:unitary_angles} 
 \end{table}

 \begin{figure}[h]
\begin{center}
\includegraphics[width=.9\columnwidth]{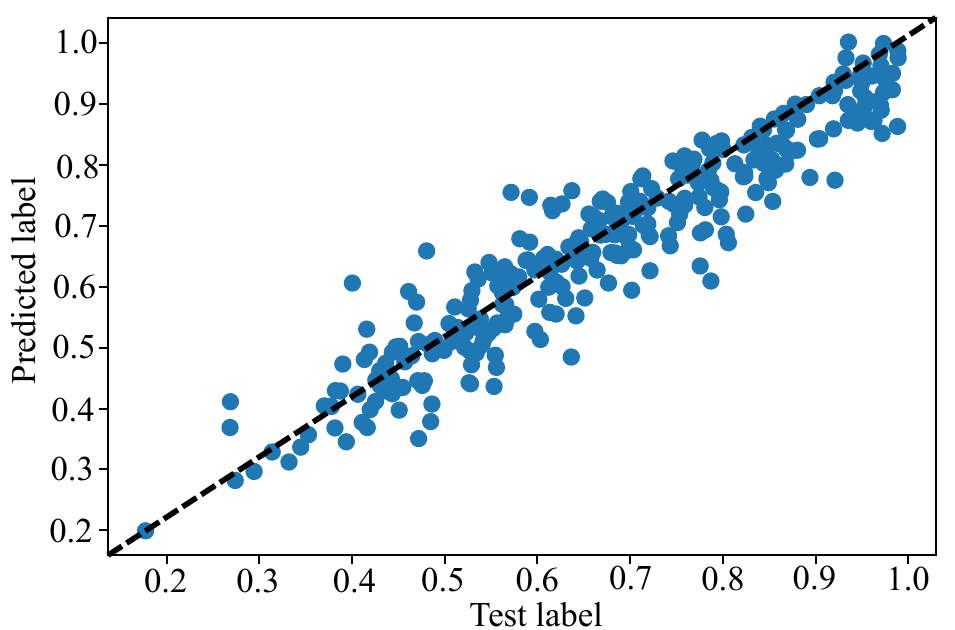}
\caption{Scatter plots for the second degree polynomial regression on the experimental dataset. The y-axis represents the labels predicted by the regression model, and the x-axis represents the actual labels. The dashed black line is the best straight line that explains the data. The $R^2$ value associated with the plot is 0.89 and the \textit{MAE} is 0.045. }
\vspace{-3mm}
\label{fig:regression_scatter_plt}
\end{center}
\end{figure}

\section{Polynomial Regression} \label{sec:4}
A regression model attempts to predict a relationship between a set of independent variables (datapoints) and an output variable (label) by utilising a polynomial function. Given a set of datapoints $\{x_i\}$, a polynomial regression model of degree $n$, finds the best prediction,  $\hat{y}$, which is an $n$-degree polynomial with input arguments $\{x_i\}$. At first, to obtain a model, one uses a part of the labeled dataset, also known as \emph{training dataset}. Once the model is obtained, to check its efficiency, one needs to employ a different group of data, known as \emph{test dataset}. Hence, a common practice is to split the training and the test set in $7{:}3$ ratio. To quantify the accuracy of the model of the dataset, we evaluate the $R^2$ value and the \emph{Mean Absolute Error (\textit{MAE})} ~\cite{hastie2009elements,bishop2013pattern}. To define these metrics, we first consider $\{y_i\}$ as our set of labels, with mean value of $\bar{y}$. We consider $\{\hat{y}_i\}$ as the predicted labels. With this, the metrics can be written as
\begin{align}
    &R^2 = 1 - \frac{\sum_i(y_i - \hat{y}_i)^2}{\sum_i(y_i - \bar{y})^2}, \nonumber \\
    &\mathit{MAE} = \frac{\sum_i|y_i-\hat{y}_i|}{N}.
\end{align}
Here, $|.|$ denotes the absolute value and $N$ is the size of the dataset. An important aspect of a predictive algorithm is to minimise \emph{overfitting}. The overfitting occurs when the model learns about the training set with so high accuracy that it fails to predict additional data. To check for the overfitting, we observe $R^2$ and \textit{MAE} score for both training data and test data. We show our results in Table~\ref{Table:different_degree}. We conclude that a polynomial regression of degree 2 achieves the least overfitting with test $R^2$ value of 0.89 and \textit{MAE} of 0.049. We show in Fig.~\ref{fig:regression_scatter_plt} the scatter plot for the second degree polynomial regression. The figure demonstrates the scatter plot between the test label and the predicted label.

\begin{table} [h]
\begin{center}
\vspace{.7cm}
\begin{tabular}{|c|c|c|c|c|}
 \hline
Deg &  Train $R^2$   & Train \textit{MAE}  \qquad &  Test $R^2$  \qquad  & Test \textit{MAE}  \qquad \\
 \hline 
 $1$ &\qquad 0.71 \qquad & \qquad 0.076 \qquad &\qquad 0.69 \qquad & \qquad 0.075 \qquad\\
  
 $2$ &\qquad 0.91 \qquad & \qquad 0.042 \qquad &\qquad 0.89 \qquad &\qquad 0.045  \qquad\\

 $3$ &\qquad 0.94 \qquad & \qquad 0.035 \qquad  &\qquad  0.85 \qquad &  \qquad 0.052 \qquad\\
 \hline
 \end{tabular}
 \caption{polynomial regression on the experimental data of size 1000. We keep $30\%$ of the experimental data as a test set and $70\%$ of the same as a training set. We vary the degree of polynomial regression (Deg). To demonstrate overfitting, we show the $R^2$ and \textit{MAE} for both training dataset and the test dataset. We observe that a polynomial regression of degree 2 achieves the least overfitting with test $R^2$ value of 0.89 and \textit{MAE} of 0.049.}
 \label{Table:different_degree} \end{center}
 \end{table}

 \emph{k-Fold Cross Validation:} A potential issue is that a  one round test-train split might result a selection bias because of the choice of test set. One way to account for it is to employ a k-fold cross validation technique \cite{K-fold}. In a k-fold cross-validation, the data-set is randomly divided into k equal sized groups. Out of the k groups, a single group is retained as the test set, and the remaining $k{-}1$ groups are the training set. Once done, in the next turn another group is selected without repetition and the entire process is iterated k-times. The results are then averaged to produce a single estimation. In our model, we use a commonly accepted value of $k{=} 10$ \cite{bishop2013pattern}. We show our results in Table~\ref{Table:K_fold}. This ensures an unbiased performance of our model. 
 
 \begin{table} [h]
\begin{tabular}{|c|c|c|}
 \hline
 Degree &  $R^2$    & \textit{MAE}   \\
 \hline 
 $1$ &\qquad 0.69${\pm}$0.07 \qquad & \qquad 0.076 $\pm$ 0.007 \qquad\\
  
 $2$ &\qquad 0.89${\pm}$0.03 \qquad &\qquad 0.045 $\pm$ 0.004 \qquad\\

 $3$ &\qquad 0.87${\pm}$0.02 \qquad &  \qquad 0.051 $\pm$ 0.004 \qquad\\
 \hline
 \end{tabular}
 \caption{k-fold cross validation on our experimental dataset for polynomial regression with degree 1,2, and 3 with value of $k$ being 10. }
 \label{Table:K_fold} 
 \end{table}

\emph{Varying the size of dataset:} It is interesting to investigate whether the algorithm performs well while training on smaller datasets. To answer this, we fix the size of the test set to $300$ and vary the length of the training set. We show our results for a second degree polynomial regression, in the Table~\ref{Table:different_percentage}. We observe that training set of size 210 achieves $R^2=0.87$ and \textit{MAE}$=0.049$. This suggests that even a small amount of experimental data is sufficient to achieve a reasonably good prediction.


\begin{table} 
\begin{center}
\vspace{.7cm}
\begin{tabular}{|c|c|c|c|c|}
 \hline
 LTD &  Train $R^2$   & Train \textit{MAE}  \qquad &  Test $R^2$  \qquad  & Test \textit{MAE}  \qquad \\
 \hline 
 $70$ &\qquad 0.86 \qquad & \qquad 0.057 \qquad &\qquad 0.15 \qquad & \qquad 0.123 \qquad\\
  
 $140$ &\qquad 0.96 \qquad & \qquad 0.031 \qquad &\qquad 0.84 \qquad &\qquad 0.053  \qquad\\

 $210$ &\qquad 0.94 \qquad & \qquad 0.036 \qquad  &\qquad  0.87 \qquad &  \qquad 0.051 \qquad\\
 
 $280$ &\qquad 0.92 \qquad & \qquad 0.040 \qquad  &\qquad 0.87 \qquad &  \qquad 0.049 \qquad\\
 
 $350$ &\qquad 0.91 \qquad & \qquad 0.042 \qquad  &\qquad 0.88 \qquad &  \qquad 0.047 \qquad\\
 
 $420$ &\qquad 0.91 \qquad & \qquad 0.043 \qquad  &\qquad 0.88 \qquad &  \qquad 0.047 \qquad\\
 
 $490$ &\qquad 0.91 \qquad & \qquad 0.043 \qquad  &\qquad 0.89 \qquad &  \qquad 0.046 \qquad\\

 $560$ &\qquad 0.91 \qquad & \qquad 0.043 \qquad  &\qquad 0.89 \qquad &  \qquad 0.046 \qquad\\
 
 $630$ &\qquad 0.91 \qquad & \qquad 0.043 \qquad  &\qquad 0.89 \qquad &  \qquad 0.046 \qquad\\
 
 $700$ &\qquad 0.91 \qquad & \qquad 0.042 \qquad  &\qquad 0.89 \qquad &  \qquad 0.045 \qquad\\
 \hline
 \end{tabular}
 \caption{Second degree polynomial trained only on the experimental data of size 1000. We keep a fixed $30\%$ of the experimental data as a test set and vary the length of training dataset (LTD). We show the $R^2$ and \textit{MAE} for both training dataset and the test dataset. We observe even with 210 training dataset, we can achieve an $R^2$ value of 0.87 and \textit{MAE} of 0.051.}
 \label{Table:different_percentage} \end{center}
 \end{table}

\section{Conclusion}
Estimating non-Markovianity can be beneficial in practical scenarios, where the environment correlates the different time-steps of a quantum experiment. We show that with only partial information about an experimental setup, we obtain a measure of non-Markovianity with fairly high accuracy. We do that by employing different machine learning models that take as input experimental data obtained through a unitary operation and state tomography. We observe that a polynomial regression model of degree 2 achieves the best performance both in terms of overfitting and performance on the test set, which is sufficiently high ($R^2{=}0.87$) even with a small number of training data (500). A high score obtained by a regression model obviates the need to employ a more intensive learning algorithm, which reduces the time-complexity of the problem. This is especially beneficial to experiments where the opportunity to collect a large dataset is limited. 
 
Our experiment is particularly interesting once we enter the large-scale quantum computation regime~\cite{Erhard_2019}.  In this regime, correlated noise among the different gates is inevitable~\cite{Correlated_error} and there is an growing interest in developing error-correcting codes for this kind of noise~\cite{Error_correction_against_correlated_noise, Corr_noise2,Corr_noise3,Corr_noise4}. Hence, our approach provides a benchmark for further noise investigation on such multi-time-step processes.

\section*{Acknowledgements}

   This work has been supported by: the Australian Research Council (ARC) by Centre of Excellence for Engineered Quantum Systems (EQUS, CE170100009). K.G. is supported by the RTP scholarship from the University of Queensland. C.G.\ is the recipient of a Sydney Quantum Academy Postdoctoral Fellowship. J.R. is supported by a Westpac Bicentennial Foundation Research Fellowship and L'Oreal-UNESCO FWIS Fellowship; F.C.\ acknowledges support through an Australian Research Council Discovery Early Career Researcher Award (DE170100712). We acknowledge the traditional owners of the land on which the University of Queensland is situated, the Turrbal and Jagera people.

\renewcommand{\theequation}{A\arabic{equation}}
\setcounter{equation}{0}

\renewcommand{\thetable}{A\arabic{table}}
\setcounter{table}{0}

\renewcommand{\thefigure}{A\arabic{figure}}
\setcounter{figure}{0}

\bibliography{Non_Markovianity.bib}

\pagebreak

\section*{Appendix} 

\emph{Mixing with Simulated data:} In practice, we may not have precise control over the environment. Hence, we ask whether assistance of simulated data augments the performance of the model. we investigate this by simulating a data set of length 14336. We proceed to vary the size of the simulated dataset and mix it with $70\%$ of the experimental dataset to train and test on the remaining $30\%$ of the experimental data. We observe that addition of simulated data deteriorates the performance of the model. To be precise, we see that the higher the number of simulated data, the worse the performance of the model. This is due to the mismatch of the experimental and simulated training data. To circumvent this, we obtain simulated data with added white noise, potentially present in the setup. We also simulate the finite sampling that occurs in the experimental procedure (we draw 50 times from a probability distribution in Eq~\ref{Eq:joint_probability}). However, we do not observe an increase in performance.
\\

\emph{Other machine learning algorithms:} It is natural to expect other conventional machine   learning algorithms might outperform the regression. We report this negatively. In this section, we demonstrate performance of several other standard machine learning algorithms, like K-Nearest Neighbour (KNN), Decision Tree, Random Forest, Support Vector Regression (SVR), and Gradient Boosting ~\cite{bishop2013pattern}. We split our experimental data into $70\%$ training set and $30\%$ test set. We show our results in Table~\ref{Table:othe_ML}. When we consider overfitting, Support Vector Regression (SVR) performs the best (test $R^2${=}0.79, train $R^2${=}0.78). Note that although Gradient boosting gives a better test $R^2$, it overfits. This suggests that polynomial regression of degree 2 is still our best choice.

\begin{table} [h]
\begin{center}
\begin{tabular}{|c|c|c|c|c|}
 \hline
Algorithm &  Train $R^2$   & Test $R^2$  \qquad & Test \textit{MAE}  \qquad \\
 \hline 
 KNN &\qquad 0.89 \qquad & 0.86 \qquad &\qquad 0.051 \qquad \\
 Decision Tree &\qquad 1.0 \qquad & 0.64 \qquad &\qquad 0.081 \qquad \\
 Random Forest &\qquad 0.98 \qquad & 0.88 \qquad &\qquad 0.049 \qquad \\
 SVR &\qquad 0.78 \qquad & 0.79 \qquad &\qquad 0.069 \qquad \\
 Gradient Boosting &\qquad 0.96 \qquad & 0.89 \qquad &\qquad 0.045 \qquad \\
   \hline
 \end{tabular}
 \caption{Different machine learning algorithms trained on the experimental data of size 1000. We split the experimental data into $30\%$ test set and $70\%$ training set and report the test and train $R^2$ and test \textit{MAE}.}
 \label{Table:othe_ML} \end{center}
 \end{table}

\end{document}